\documentstyle[11pt]{article}
\begin{document}

%%%%%%%%%%%%%%%%%%%%%%%%%%%%%Start of Text%%%%%%%%%%%%%%%%%%%%%%%%%%%%%%%%%
\bigskip
{\Large\bf
\centerline{$B\rightarrow X_s\tau^+\tau^-$ in a two Higgs doublet model}
\bigskip
\normalsize

\centerline{Yuan-Ben Dai$^1$,~~~~Chao-Shang Huang$^1$,~~~~Han-Wen Huang$^{2,1}$}
\centerline{\sl $^1$ Institute of Theoretical Physics, Academia Sinica,
      P.O.Box 2735,}
\centerline{\sl Beijing 100080, P.R.China}
\centerline{\sl $^2$ CCAST (World Laboratory), Beijing 100080, P.R.China} 
\bigskip

\begin{abstract}
\bf The inclusive rate and forward-backward asymmetry of dilepton angular 
distribution for a B-meson to decay to strange hadronic final states and a
$\tau^+\tau^-$ pair in a two Higgs doublet model are computed. In particular,
contributions of neutral Higgs bosons to the decay are included. QCD corrections
to the effective Hamiltonian for $B\rightarrow X_s\tau^+\tau^-$ are calculated
using the leading logarithmic approximation.
\end{abstract}

\vfill\eject\pagestyle{plain}\setcounter{page}{1}

Flavor changing neutral current (FCNC) transitions $B\rightarrow X_s\gamma$ and
$B\rightarrow X_sl^+l^-$ provide testing grounds for the standard model (SM) at
the loop level and sensitivity to new physics. Rare decays $B\rightarrow X_sl^+l^-(l=e,\mu)$ have been extensively investigated in both SM and the two Higgs 
doublet models (2HDM) [1-13]. In these processes contributions from exchanging neutral
Higgs bosons can be safely neglected because of smallness of $\frac{m_l}{m_W}
(l=e,\mu)$. Recently, $\tau$ polarization asymmetry in $B\rightarrow X_s\tau^+
\tau^-$ in SM has been studied \cite{14} by using the method same as that for 
$B\rightarrow X_sl^+l^-(l=e,\mu)$. The author of ref.\cite{14} did not
consider the role played by the neutral Higgs boson because the contributions
due to exchanging the neutral Higgs boson in SM can also be neglected, comparing
with those due to $\gamma,~Z$.

In this note we investigate the inclusive decay $B\rightarrow X_s\tau^+\tau^-$
in a 2HDM. We consider the 2HDM in which the up-type quarks get masses from
Yukawa couplings to the one Higgs doublet $H_2$ (with the vacuum expectation value $v_2$) and down-type quarks and leptons get masses from Yukawa 
couplings to the another Higgs doublet $H_1$ (with the vacuum expectation value
$v_1$). Such a model occurs as a natural feature in supersymmetric theories.
The Higgs boson couplings to quarks and leptons depend on the ratio $tg\beta
=\frac{v_2}{v_1}$ which is a free parameter in the model. Constraints on
$tg\beta$ from $K-\bar{K}$ and $B-\bar{B}$ mixing, $\Gamma(b\rightarrow s\gamma)
$,$\Gamma(b\rightarrow c\tau\bar{\nu}_{\tau})$ and $R_b$ have been given \cite{15,16}
\begin{equation}
0.7\le tg\beta\le 0.6(\frac{m_{H^{\pm}}}{1Gev})
\end{equation}
(and the lower limit $m_{H^{\pm}}\ge 200Gev$ has also been given in the ref.
\cite{16}). It is obvious that the contributions from exchanging neutral Higgs 
bosons now is enhanced roughly by a factor of $tg^2\beta$ and can compete with
those from exchanging $\gamma,~Z$ when $tg\beta$ is large enough.

Inclusive decay rates of heavy hadrons can be calculated in heavy quark 
effective theory (HQET) \cite{17} and it has been shown that the leading terms in $1/m_Q$
expansion turns out to be the decay of a free (heavy) quark and corrections stem
from the order $1/m_Q^2$ \cite{18}. In what follows we shall calculate the leading term (the
$1/m^2_Q$ correction can be easily added if needed). The transition rate for
$b\rightarrow s\tau^+\tau^-$ can be computed in the framework of the QCD 
corrected effective weak hamiltonian, obtained by integrating out the top quark,
Higgs bosons and $W^{\pm},Z$ bosons
\begin{equation}\label{ham}
H_{eff}=\frac{4G_F}{\sqrt{2}}V_{tb}V^*_{ts}(\sum_{i=1}^{10}C_i(\mu)O_i(\mu)
+\sum_{i=1}^{10}C_{Q_i}(\mu)Q_i(\mu))
\end{equation}
where $O_i(i=1,\cdots ,10)$ is the same as that given in the ref.\cite{4}, $Q_i$'s
come from exchanging the neutral Higgs bosons and are defined by
\begin{eqnarray}\nonumber
Q_1&=&\frac{e^2}{16\pi^2}(\bar{s}^{\alpha}_Lb^{\alpha}_R)(\bar{\tau}
\tau)\\\nonumber
Q_2&=&\frac{e^2}{16\pi^2}(\bar{s}^{\alpha}_Lb^{\alpha}_R)(\bar{\tau}\gamma_5
\tau)\\\nonumber
Q_3&=&\frac{g^2}{16\pi^2}(\bar{s}^{\alpha}_Lb^{\alpha}_R)(\sum_q\bar{q}^{\beta}
_Lq^{\beta}_R)\\\nonumber
Q_4&=&\frac{g^2}{16\pi^2}(\bar{s}^{\alpha}_Lb^{\alpha}_R)(\sum_q\bar{q}^{\beta}
_Rq^{\beta}_L)\\\nonumber
Q_5&=&\frac{g^2}{16\pi^2}(\bar{s}^{\alpha}_Lb^{\beta}_R)(\sum_q\bar{q}^{\beta}
_Lq^{\alpha}_R)\\\nonumber
Q_6&=&\frac{g^2}{16\pi^2}(\bar{s}^{\alpha}_Lb^{\beta}_R)(\sum_q\bar{q}^{\beta}
_Rq^{\alpha}_L)\\\nonumber
Q_7&=&\frac{g^2}{16\pi^2}(\bar{s}^{\alpha}_L\sigma^{\mu\nu}b^{\alpha}_R)
(\sum_q\bar{q}^{\beta}_L\sigma_{\mu\nu}q^{\beta}_R)\\\nonumber
Q_8&=&\frac{g^2}{16\pi^2}(\bar{s}^{\alpha}_L\sigma^{\mu\nu}b^{\alpha}_R)
(\sum_q\bar{q}^{\beta}_R\sigma_{\mu\nu}q^{\beta}_L)\\\nonumber
Q_9&=&\frac{g^2}{16\pi^2}(\bar{s}^{\alpha}_L\sigma^{\mu\nu}b^{\beta}_R)
(\sum_q\bar{q}^{\beta}_L\sigma_{\mu\nu}q^{\alpha}_R)\\
Q_{10}&=&\frac{g^2}{16\pi^2}(\bar{s}^{\alpha}_L\sigma^{\mu\nu}b^{\beta}_R)
(\sum_q\bar{q}^{\beta}_R\sigma_{\mu\nu}q^{\alpha}_L)
\end{eqnarray}
with $e$ and $g$ being the electromagnetic and strong coupling constants 
respectively.

At the renormalization point $\mu=m_W$ the coefficients $C_i$'s in the 
effective hamiltonian have been given in the ref.\cite{4} and $C_{Q_i}$'s are
(neglecting the $O(tg\beta)$ term)\footnote{We use the Feynman rules given in 
ref.\cite{19}.}.
\begin{eqnarray}\nonumber
C_{Q_1}(m_W)&=&\frac{m_bm_{\tau}}{m^2_{h^0}}tg^2\beta\frac{1}{sin^2\theta_W}
\frac{x}{4}\{(sin^2\alpha+hcos^2\alpha)f_1(x,y)\\\nonumber
&+&[m^2_{h^0}/m_w^2+(sin^2\alpha+hcos^2\alpha)(1-z)]f_2(x,y)\\
&+&\frac{sin^22\alpha}
{2m^2_{H^{\pm}}}(m^2_{h^0}-\frac{(m^2_{h^0}+m^2_{H^0})^2}{2m^2_{H^0}})f_3(y)\},\\
C_{Q_2}(m_W)&=&-\frac{m_bm_{\tau}}{m^2_{A^0}}tg^2\beta\{f_1(x,y)+(1+
\frac{m^2_{H^{\pm}}-m^2_{A^0}}{m^2_W})f_2(x,y)\},\\
C_{Q_3}(m_W)&=&\frac{m_be^2}{m_{\tau}g^2}(C_{Q_1}(m_W)+C_{Q_2}(m_W)),\\
C_{Q_4}(m_W)&=&\frac{m_be^2}{m_{\tau}g^2}(C_{Q_1}(m_W)-C_{Q_2}(m_W)),\\
C_{Q_i}(m_W)&=&0, ~~~~i=5,\cdots, 10
\end{eqnarray}
where
$$
x=\frac{m^2_t}{m^2_W},~~y=\frac{m^2_t}{m^2_{H^{\pm}}},
$$
$$
z=\frac{x}{y},~~h=\frac{m^2_{h^0}}{m^2_{H^0}},
$$
$$
f_1(x,y)=\frac{xlnx}{x-1}-\frac{ylny}{y-1},
$$
$$
f_2(x,y)=\frac{xlny}{(z-x)(x-1)}+\frac{lnz}{(z-1)(x-1)},
$$
$$
f_3(y)=\frac{1-y+ylny}{(y-1)^2}.
$$
Neglecting the strange quark mass, the effective hamiltonian (\ref{ham}) leads 
to the following matrix element for $b\rightarrow s\tau^+\tau^-$
\begin{eqnarray}\label{matrix}\nonumber
M&=&\frac{G_F\alpha}{\sqrt{2}\pi}V_{tb}V^*_{ts}[C^{eff}_8\bar{s}_L\gamma_{\mu}
b_L\bar{\tau}\gamma^{\mu}\tau+C_9\bar{s}_L\gamma_{\mu}b_L\bar{\tau}\gamma^{\mu}
\gamma^5\tau\\
&+&2C_7m_b\bar{s}_Li\sigma^{\mu\nu}\frac{q^{\nu}}{q^2}b_R\bar{\tau}\gamma^{\mu}
\tau+C_{Q_1}\bar{s}_Lb_R\bar{\tau}\tau+C_{Q_2}\bar{s}_Lb_R\bar{\tau}\gamma^5
\tau],
\end{eqnarray}
where
\begin{eqnarray}\label{coeff}\nonumber
C^{eff}_8&=&C_8+\{g(\frac{m_c}{m_b},\hat{s})\\
&+&\frac{3}{\alpha^2}k\sum_{V_i=
\psi^{\prime}}\frac{\pi M_{V_i}\Gamma(V_i\rightarrow\tau^+\tau^-)}{M^2_{V_i}-q^2
-iM_{V_i}\Gamma_{V_i}}\}(3C_1+C_2),
\end{eqnarray}
with $\hat{s}=q^2/m_b^2,~~q=(p_{\tau^+}+p_{\tau^-})^2$. In (\ref{coeff}) 
$g(\frac{m_c}{m_b},\hat{s})$ arises from the one-loop matrix element of the four-quark 
operators $O_i$'s and has been given in \cite{4}. The final term in (10)
estimates the long-distance contribution from the intermediate $\psi^{\prime}$
\cite{10}.

The QCD corrections to coefficients $C_i$ and $C_{Q_i}$ can be incooperated
in the standard way by using the renormalization group equations. The mixing 
of the operators $O_i(i=1,2,\cdots,10)$ at $\alpha_s$ order has been studied 
and the corresponding anomalous dimension matrix has been given \cite{4,13}. 
(The mixing of $O_i$ at the next-to-leading order has also been studied in 
ref. \cite{13}.) We studied the one-loop mixing of the new set of operators
${Q_i}$ and found that the corrections to coefficients due to mixing with
$Q_i$ are small and can be neglected provided that $tg\beta<50$. For example,
including the mixing between $O_7$ and $Q_3$ (and neglecting the mixing of the
other $Q_i$ with $O_7$ and the mixing of $Q_3$ with the other $Q_i$), we have
\begin{equation}\label{c70}
C_7(m_b)=\eta^{-16/23}\{C_7(m_W)-[\frac{58}{135}(\eta^{10/23}-1)+\frac{29}{189}
(\eta^{28/23}-1)]C_2(m_W)-\frac{1}{34}(\eta^{17/23}-1)C_{Q_3}(m_W)\}
\end{equation}
where $\eta=\alpha_s(m_b)/\alpha_s(m_W)$. The additional term is less than $5\%$
correction to the value of $C_7(m_b)$ provided $tg\beta<50$. If we include the 
mixing of $Q_i(i=3,5,7,9)$ with $O_7$
\begin{equation}
\gamma^{(0)}=\begin{tabular}{c|c}
&$O_7$\\\hline$Q_3$&1/6\\$Q_5$&1/2\\$Q_7$&-1/6\\$Q_9$&-1/2
\end{tabular}
\end{equation}
(and neglect the mixing of $Q_i(i=4,6,8,10)$ with $O_7$) and also consider the 
mixing among $Q_3,Q_5,Q_7,Q_9$ which can be derived from Eq.(39) of
 ref.\cite{20} and is given by
\begin{equation}
\gamma^{(0)}=\begin{tabular}{c|cccc}
&$Q_3$&$Q_5$&$Q_7$&$Q_9$\\\hline
$Q_3$&$-8+\beta_0$&0&2/3&-2\\
$Q_5$&-3&$1+\beta_0$&-1&-7/3\\
$Q_7$&2&-6&$8/3+\beta_0$&0\\
$Q_9$&-3&-7&0&$-19/3+\beta_0$
\end{tabular}
\end{equation}
with $\beta_0=11-\frac{2}{3}n_f$, then we have 
\begin{equation}\label{c7}
C_7(m_b)=\eta^{-16/23}\{C_7(m_W)-[\frac{58}{135}(\eta^{10/23}-1)
+\frac{29}{189}(\eta^{28/23}-1)]C_2(m_W)-0.012C_{Q_3}(m_W)\}
\end{equation}
The final term in (\ref{c7}) is still small ($3\%$ correction) when $tg\beta
<50$. Using (\ref{c70}) or (\ref{c7}), the limit on $tg\beta$ obtained 
from $\Gamma(b\rightarrow s\gamma)$ is $\frac{tg\beta}{m_{A^0}}<0.25Gev^{-1}$.\\ 

$Q_i(i=1,\cdots,10)$ does not mix with $O_8,O_9$ so that $C_8$ and $C_9$
remain unchanged and are given in ref.\cite{4}
\begin{eqnarray}
C_8(m_b)&=&C_8(m_W)+\frac{4\pi}{\alpha_s(m_W)}[-\frac{4}{33}(1-\eta^{-11/23})
+\frac{8}{87}(1-\eta^{-29/23})]C_2(m_W),\\
C_9(m_b)&=&C_9(m_W).
\end{eqnarray}

It is obvious that operators $O_i(i=1,\cdots,10)$ and $Q_i(i=3,\cdots,10)$
do not mix into $Q_1$ and $Q_2$ and also there is no mixing between $Q_1$
and $Q_2$. Therefore, the evolution of $C_{Q_1},C_{Q_2}$ is controlled by the 
anomalous dimensions of $Q_1,Q_2$ respectively.
\begin{equation}
C_{Q_i}(m_b)=\eta^{-\gamma_Q/\beta_0}C_{Q_i}(m_W),~~i=1,2,
\end{equation}
where $\gamma_Q=-4$ \cite{21} is the anomalous dimension of $\bar{s}_Lb_R$.

From Eq.(\ref{matrix}), it is easy to derive the double differential distribution
$\frac{d^2\Gamma}{d\hat{s}dcos\theta}$ as follows
\begin{eqnarray}\label{width}\nonumber
\frac{d^2\Gamma(B\rightarrow X_s\tau^+\tau^-)}{d\hat{s}dz}&=&B(B\rightarrow
X_cl\bar{\nu})\frac{\alpha^2}{4\pi^2f(\frac{m_c}{m_b})}(1-\hat{s})^2
(1-\frac{4t^2}{\hat{s}})^{\frac{1}{2}}\frac{|V_{tb}V_{ts}^*|^2}
{|V_{cb}|^2}\\\nonumber
&&\{\frac{3}{2}|C^{eff}_8|^2[(1+\hat{s})-(1-\hat{s})(1-\frac{4t^2}{\hat{s}})z^2
+4t^2]\\\nonumber
&+&6|C_7|^2[(1+\frac{1}{\hat{s}})-(1-\frac{4t^2}{\hat{s}})(1-\frac{1}{\hat{s}})
z^2+\frac{4t^2}{\hat{s}^2}]\\\nonumber
&+&\frac{3}{2}|C_9|^2[(1+\hat{s})-(1-\hat{s})(1-\frac{4t^2}{\hat{s}})z^2-4t^2]
\\\nonumber
&+&12Re(C^{eff}_8C^*_7)(1+\frac{2t^2}{\hat{s}})
+6Re(C_8C_9^*)(1-\frac{4t^2}{\hat{s}})^{\frac{1}{2}}\hat{s}z\\\nonumber
&+&12Re(C_7C_9^*)(1-\frac{4t^2}{\hat{s}})^{\frac{1}{2}}z+\frac{3}{2}|C_{Q_1}|^2
(\hat{s}-4t^2)+\frac{3}{2}|C_{Q_2}|^2\hat{s}\\\nonumber
&+&6Re(C_8C_{Q_1}^*)(1-\frac{4t^2}{\hat{s}})^{\frac{1}{2}}tz
+12Re(C_7C_{Q_1}^*)(1-\frac{4t^2}{\hat{s}})^{\frac{1}{2}}tz\\
&+&6Re(C_9C_{Q_2}^*)t\},
\end{eqnarray}
where $z=cos\theta,~~t=\frac{m_{\tau}}{m_b}$ and $\theta$ is the angle between 
the momentum of the B-meson and that of $l^+$ in the center of mass frame of the
dileptons $\tau^+\tau^-$. Integrating the angle variable, we obtain the 
invariant dilepton mass distribution
\begin{eqnarray}\nonumber
\frac{d\Gamma(B\rightarrow X_s\tau^+\tau^-)}{d\hat{s}}&=&B(B\rightarrow
X_cl\bar{\nu})\frac{\alpha^2}{4\pi^2f(\frac{m_c}{m_b})}(1-\hat{s})^2
(1-\frac{4t^2}{\hat{s}})^{\frac{1}{2}}\frac{|V_{tb}V_{ts}^*|^2}
{|V_{cb}|^2}\\\nonumber
&&\{|C^8_{eff}|^2(1+\frac{2t^2}{\hat{s}})(1+2\hat{s})+4|C_7|^2 
(1+\frac{2t^2}{\hat{s}})(1+\frac{2}{\hat{s}})\\\nonumber
&+&|C_9|^2[(1+2\hat{s})+\frac{2t^2}{\hat{s}}(1-4\hat{s})]
+12Re(C_7C^{8*}_{eff})(1+\frac{2t^2}{\hat{s}})\\
&+&\frac{3}{2}|C_{Q_1}|^2(\hat{s}-4t^2)+\frac{3}{2}|C_{Q_2}|^2\hat{s}
+6Re(C_9C_{Q_2}^*)t\}.
\end{eqnarray}
Switching off the channel exchanging neutral Higgs bosons, Eq.(\ref{width}) turns
out to be the same as that given in Ref.[14]. We also give the forward-backward
asymmetry in $B\rightarrow X_s\tau^+\tau^-$
\begin{eqnarray}\nonumber
A(\hat{s})&=&\int_0^1dz\frac{d^2\Gamma}{d\hat{s}dz}-\int_{-1}^0dz
\frac{d^2\Gamma}{d\hat{s}dz}\\\nonumber
&=&B(B\rightarrow X_cl\bar{\nu})\frac{3\alpha^2}{2\pi^2f(\frac{m_c}{m_b})}
(1-\hat{s})^2(1-\frac{4t^2}{\hat{s}})\frac{|V_{tb}V_{ts}^*|^2}{|V_{cb}|^2}\\
&&\{Re(C_8C_9^*)\hat{s}+2Re(C_7C_9^*)+Re(C_8C_{Q_1}^*)t+2Re(C_7C_{Q_1}^*)t\}.
\end{eqnarray}

The following parameters have been used in the numerical calculations:
$$
m_t=175Gev,~m_b=5.0Gev,~m_c=1.6Gev,~m_{\tau}=1.7Gev,~\Lambda_{\overline{MS}}=0.1Gev.
$$
Numerical results are summarized in Fig.1-2.

\begin{figure}
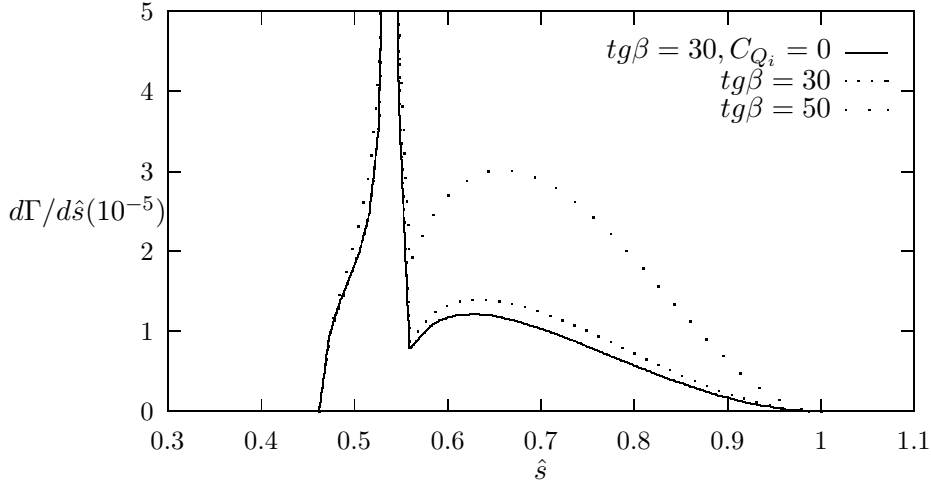

\input fig1.tex
\caption{The differential branching ratio $\frac{d\Gamma}{d\hat{s}}$ (normalized 
to $\Gamma(B\rightarrow X_ce\hat{\nu}_e)$) as a function of the scaled invariant 
dilepton mass $\hat{s}=s/m_b^2$ in the decay $B\rightarrow X_s\tau^+\tau^-$. 
We have taken $m_{H^{\pm}}=200Gev$, $m_{h^0}=80Gev$, $m_{H^0}=150Gev$, 
and $m_{A^0}=100Gev$. Assumed $tg\beta=30,50$ are indicated on the curves. The 
solid line corresponds to the case of switching off the channel exchanging the neutral Higgs bosons.}
\end{figure}

\begin{figure}
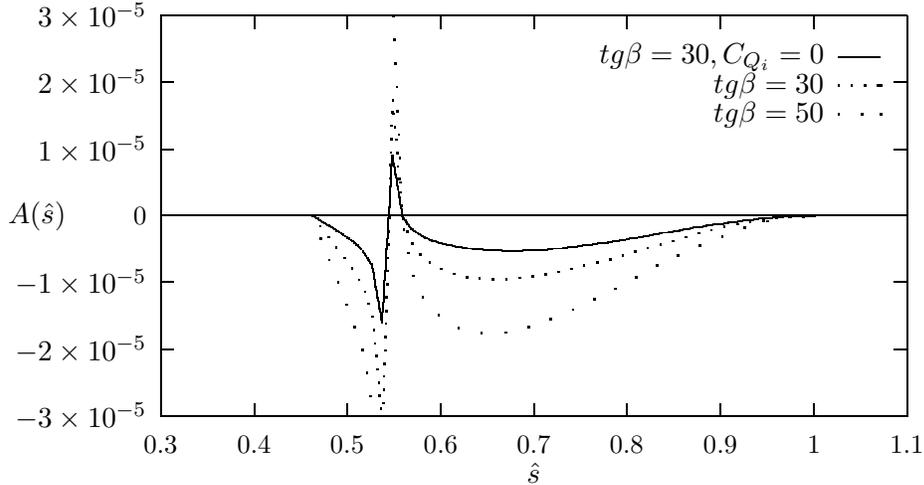

\input fig2.tex
\caption
{Forward-backward asymmetry of the dileptons on the decay 
$B\rightarrow X_s\tau^+\tau^-$, $A(\hat{s})$ (normalized to 
$\Gamma(B\rightarrow X_ce\hat{\nu}_e)$), as a function of the scaled invariant 
dilepton mass $\hat{s}=s/m_b^2$. We have taken $m_{H^{\pm}}=200Gev$, 
$m_{h^0}=80Gev$, $m_{H^0}=150Gev$, and $m_{A^0}=100Gev$. Assumed 
$tg\beta=30,50$ are indicated on the curves.  The solid line corresponds to the case of switching off the channel exchanging the neutral Higgs bosons.}
\end{figure}

It is obvious that there are sharp peaks on the curves of differential
branching ratio $\frac{d\Gamma}{d\hat{s}}$ and asymmetry $A(\hat{s})$ around
the mass of resonance $\psi^{\prime}$. One can see from the figures. that the contributions of
neutral Higgs bosons is significant in 2HDM
when  $tg\beta$ is larger than 25 and the values of the masses of the Higgs bosons are in the reasonable range, and the forward-backward asymmetry of dilepton angular distribution is more sensitive to $tg\beta$ than the invariant mass distribution. We expect that the more precise experiments on rare decays of B mesons
will shed light on the two Higgs doublet model.

\vspace{5mm}
This work is supported in part by the National Natural Science Fundation of China and the Postdoctoral Science Foundation of China.

\vfill\eject

%%%%%%%%%%%%%%%%%%%%%%%%%%%% Figure Caption %%%%%%%%%%%%%%%%%%%%%%%%%%%%%%%%%%


\begin{thebibliography}{99}
\bibitem{1}{W.-S.Hou, R.S.Willey and A.Soni, {\it Phys.
Rev. Lett}{\bf 58}(1987)1608.}
\bibitem{2}{N.G.Deshpande and J.Trampetic, {\it Phys.Rev.Lett.}{\bf 60}
(1988)2583.}
\bibitem{3}{C.S.Lim, T.Morozumi and A.I.Sanda, {\it Phys.Lett.}{\bf B218}
(1989)343.}
\bibitem{4}{B.Grinstein, M.J.Savage and M.B.Wise, {\it Nucl.Phys.}{\bf B319}
(1989)271.}
\bibitem{5}{C.Domingues, N.Paver and Riazuddin, {\it Phys.Lett.}{\bf B214}
(1988)459.}
\bibitem{6}{N.G.Deshpande, J.Trampetic and K.Ponose,{\it Phys.Rev.}{\bf D39}
(1989)1461.}
\bibitem{7}{W.Jaus and D.Wyler, {\it Phys.Rev.}{\bf D41}(1990)3405.}
\bibitem{8}{P.J.O'Donnell and H.K.K.Tung,{it Phys.Rev.}{\bf D43}(1991)2067.}
\bibitem{9}{N.Paver and Riazuddin, ICTP Trieste report(1991).}
\bibitem{10}{A.Ali, T.Mannel and T.Morozumi, {\it Phys.Lett.}{\bf B273}
(1991)505.}
\bibitem{11}{A.Ali, G.F.Giudice and T.Mannel, {\it Z.Phys.}{\bf C67}(1995)417}
\bibitem{12}{C.Greub, A.Ioannissian and D.Wyler,{\it Phys.Lett.}{\bf B346}
(1995)149; D.Liu, Univ. of Tasmania Report(1995); G.Burdman, Fermilab Report 
Pub-95/113-T(1995).}
\bibitem{13}{A.J.Buras and M.M\"{u}nz,{\it Phys.Rev.}{\bf D52}(1995)186.}
\bibitem{14}{JoAnne L.Hewett, SLAC-PUB-95-6820.}
\bibitem{15}{J.Kalinowski,{\it Phys.Lett.}{\bf B245}(1990)201.}
\bibitem{16}{A.K.Grant,{\it Phys.Rev.}{\bf D51}(1995)207.}
\bibitem{17}{For a comprehensive review, see: M.Neubert,{\it Phys.Rep.}
{\bf 245}(1994)396.}
\bibitem{18}{I.I.Bigi, M.Shifman, N.G.Vraltsev and A.I.Vainshtein, {\it Phys.
Rev.Lett.}{\bf 71}(1993)496; B.Blok, L.Kozrakh, M.Shifman and A.I.Vainshtein,
{\it Phys.Rev.}{\bf D49}(1994)3356; A.V.Manohar and M.B.Wise, {\it Phys.Rev.}
{\bf D49}(1994)1310; S.Balk, T.G.K\"orner, D.Pirjol and K.Schilcher,
{\it Z.Phys.} {\bf C64}(1994)37; A.F.Falk, Z.Ligeti, M.Neubert and Y.Nir,
{\it Phys.Lett.}{\bf B326}(1994)145.}
\bibitem{19}{J.F.Gunion,H.E.Haber,G.Kane and S.Dawson, {\it The Higgs hunter's
guide} (Addison-Wesley, Reading, MA, 1990).}
\bibitem{20}{H.Anlau, SLAC-PUB-6525(1994).}
\bibitem{21}{C.S.Huang, {\it Commun. Theor. Phys.} {\bf 2}(1983)1265.}
\end{thebibliography}
\end{document}